\documentclass[twocolumn]{article}

\usepackage[margin=2cm]{geometry}
\usepackage{titlesec}
\usepackage{graphicx}
\usepackage[font=footnotesize, figurename=\textsc{Figure}]{caption}
\usepackage{amsmath}
\usepackage{booktabs}
\usepackage{authblk}
\usepackage[style=ieee,backend=biber]{biblatex}

\titleformat{\section}{\normalfont\bfseries\filcenter\MakeUppercase}{\thesection.}{1em}{}

\renewcommand{\thesection}{\Roman{section}}

\graphicspath{{figures/}}

\bibliography{biblio_letter.bib}

\begin{document}
	\title{\Large{\textbf{RCS angular control with gradient metasurfaces:\\ design and measurement}}}
	\date{}

	\author[1,2,*]{Matthieu Elineau}
	\author[1]{Renaud Loison}
	\author[1]{St\'ephane M\'eric}
	\author[1]{Raphaël Gillard}
	\author[2]{Pascal Pagani}
	\author[2]{Genevi\`eve~Maz\'e-Merceur}
	\author[3]{Philippe Pouliguen}
	\affil[1]{Univ-Rennes, INSA Rennes, CNRS, IETR-UMR 6164, F-35 000 Rennes, France}
	\affil[2]{CEA CESTA, Le Barp, France}
	\affil[3]{Defense Innovation Agency, Ministry of the Armed Forces, Paris, France}
	\affil[*]{e-mail: matthieu.elineau@proton.me}

	\renewcommand\Authfont{\fontsize{11}{13.2}\selectfont}
	\renewcommand\Affilfont{\fontsize{9}{10.8}\selectfont}

	\renewcommand{\abstractname}{\vspace{-3\baselineskip}}
	\twocolumn[
	\begin{@twocolumnfalse}
		\maketitle
		\begin{abstract}
			This letter proposes the design and measurement of a periodic metasurface that achieves anomalous reflection with reduced RCS in a given parasitic direction. A previous study proposed a semi-analytical model to predict the RCS behavior of such a metasurface. However, this first study did not include any experimental exploration to verify the theoretical results. To complete this study, this work presents an experimental validation of the proposed design, with a focus on manufacturing and measurement issues. The synthesis, design specifications, fabrication method and experimental setup are presented and discussed. Measurement results are also examined in detail, highlighting some limitations in metasurfaces RCS measurements. The proposed metasurface effectively achieves the predicted RCS level reduction in the considered parasitic direction. The agreement between simulation and experimental results demonstrates the accuracy of the modelling and the efficiency of the optimisation procedure.\\
		\end{abstract}
	\end{@twocolumnfalse}
	]

\section{Introduction}
	Radar cross-section (RCS) is a measure of a target’s reflectivity~\cite{Knott2004}. For stealth purposes, we may want to reduce the RCS of an object. The classic method is to use shape concepts or volume-absorbing materials. Another type of material has recently emerged: metasurfaces. They have generated a wide variety of applications for wave front shaping~\cite{Glybovski2016}. Among them, the classical gradient metasurface leading to anomalous reflection has been particularly studied~\cite{Yu2011}. It allows reflecting an incoming wave in a non-specular direction, which may prove useful for stealth purposes. Such a metasurface consists of the juxtaposition of sub-wavelength scattering elements (the cells) introducing a linearly varying phase shift at the surface of the object. However, these classical gradient metasurfaces exhibit parasitic reflection directions because of the periodic nature of the phase gradient~\cite{DiazRubio2017}. Such parasitic reflections need to be mitigated in order to achieve accurate RCS angular control. The scatterers used to cover a gradient period form a supercell that, when periodically replicated to pave the overall surface, excites Floquet harmonics~\cite{Bhattacharyya2006} that lead to parasitic reflections. Such translational invariances and associated symmetries in metasurfaces are now widely studied~\cite{QuevedoTeruel2020}.

	Recently, we proposed a method~\cite{Elineau2022a} to mitigate parasitic reflections of an anomalous reflecting metasurface with a monodimensionally varying gradient, using Floquet analysis. Simulating a supercell in a Floquet environment allows the computation of the $S_{m, n}$ scattering parameters between the different Floquet modes. This is a way to take into account coupling effects that result in changes in the phase responses of individually characterized cells. Assuming $n = 0$ corresponds to the fundamental mode impinging on the metasurface, there is a direct correlation between $S_{m, 0}$ and the resulting RCS level in the propagation direction of reflected mode $m$, defined by an angle of incidence $\theta_m$. This means that the outputs of the Floquet simulation (light to simulate) may be sufficient to correctly describe the behavior of the whole structure (requiring a much heavier simulation) in terms of RCS. Going further in this direction, the Floquet simulation $S$ parameters have been incorporated into an analytical model to calculate the total RCS of the structures. This approach has been presented in~\cite{Elineau2022b} proving that the outputs of the Floquet type simulation are indeed sufficient to estimate the RCS of the metasurfaces. A RCS control scenario was proposed where the incoming radiation is accurately redirected into a selected direction, while maintaining low RCS into the parasitic direction. The purpose of the present letter is to conclude the demonstration by providing the design and measurement of such a metasurface and, more generally speaking, to address the associated practical fabrication and measurement issues. It also provides a wideband characterization of the optimized metasurface, for a design optimisation at a single frequency. This characterization outlines important considerations for the use of such resonant metasurfaces in stealth applications, indicating that achieving broadband performance may require design strategies beyond the simple monolayer approach proposed here.

\section{Metasurfaces synthesis}
	The complete motivation of the design process, as well as the synthesis and optimisation of the metasurface are explained in detail in~\cite{Elineau2022b}, and we recall the main steps hereafter for the sake of completeness. The metasurface initial objective is to reflect the most part of a normally incident plane wave into the $\theta_1 = 60^\circ$ direction, with a classical gradient metasurface. The wave reflection occurs in the plane normal to the surface and containing the direction of the gradient. The incoming plane wave is TM polarised and the working frequency is 8~GHz.
	
\begin{figure}[]
		\centering
		\includegraphics[width=\linewidth]{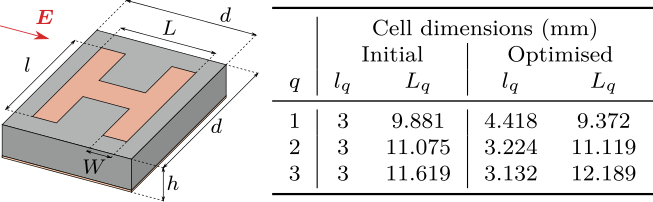}
		\caption{The H shape phase shifting cell. Fixed parameters of the cell are $\epsilon_r = 2.17$, $h = 1.6$~mm, $d = 14.43$~mm and $W = 2$~mm. The $l$ and $L$ dimensions are used as degrees of freedom to produce the required values for the phase gradient.}\label{fig_Hcell}
	\end{figure}

	The initial surface is created by designing individually three cells that produce the linear phase variation required along a gradient period, constituting the supercell. In the initial case, the three cells are simply chosen to be 120$^\circ$ apart in terms of phase response so the full gradient period is covered. One of them is set to be at a 0$^\circ$ phase. The number of three cells is the lowest number of cells in this situation were each cell is guaranteed to be smaller than $\frac{\lambda}{2}$. Each cell phase response is simulated using local periodicity assumption, under normal illumination. Cells dimensions corresponding to the required phases are found in the table of Figure~\ref{fig_Hcell}, with the description of the cell geometrical parameters. The surface is created by replicating the supercell nine times to create an array that is approximately $10\lambda$ long. Figure~\ref{fig_sim_mes} shows the simulated RCS of the initial surface, in blue dashed line. The surface mainly radiates in the intended $\theta_1 = 60^\circ$ direction but also shows a quite high parasitic reflection level in the Floquet direction $\theta_{-1} = -60^\circ$.

	In order to reduce the parasitic reflection, an optimised surface is proposed. The optimisation is performed at a unique frequency $f = 8$~GHz and consists of the maximisation of the difference between the $S_{1,0}$ and $S_{-1,0}$ parameters of the supercell, using cell dimensions as degrees of freedom. A gradient type optimization is used which starts from the initial supercell geometry. The $L_q$ dimensions are firstly used as degrees of freedom as they define the length of the parts of the patch that are aligned with the electric field, providing stronger phase variations. The phase responses are then more finely tuned with the use of $l_q$ as degrees of freedom. This process gives an optimised supercell whose dimensions are also listed in Figure~\ref{fig_Hcell}. In general, the optimisation process makes cells move away from resonance. In order to evaluate RCS reductions we define the figure of merit $\delta\mathrm{RCS} = \frac{\mathrm{RCS}(-60^\circ)}{\mathrm{RCS}(60^\circ)}$ so reductions are evaluated relatively to the maximum radiation direction. As shown in Figure~\ref{fig_sim_mes}, $\delta\mathrm{RCS} = -8.8$~dB for the initial simulated case (dashed line) while for the optimised simulated case (solid line), $\delta\mathrm{RCS} = -16.2$~dB which results in a $7.4$~dB reduction of the parasitic lobe level, relatively to the main lobe level.

\section{Metasurfaces manufacturing}
\begin{figure}[]
		\centering
		\includegraphics[width=\linewidth]{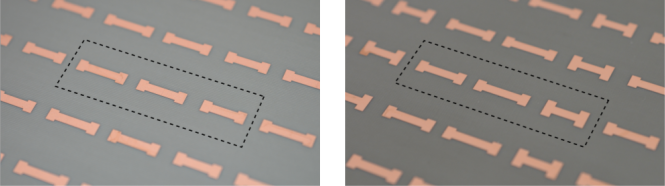}
		\caption{Close view of the fabricated metasurfaces. Initial (left) and optimised (right) surfaces. A supercell is represented with dashed lines.}\label{fig_fab_surf}
	\end{figure}

	Both surfaces, initial and optimised are fabricated with printed copper patches on a Neltec NY9217 copper backed substrate of thickness $h = 1.6$~mm. The relative permittivity of such a substrate is low (${\epsilon_r = 2.17}$) which provides smooth phase variations of the cells with their geometrical parameter as well as with $\theta$ variations. The impact of low permittivity values are discussed in~\cite{Kedze2022}. This helps under two aspects, at least. On one hand, finding converging optimisations is made easier and, on the other hand, measurements are made less sensitive to uncertainties about the angle $\theta$. A 27 $\times$ 27 cells surface is fabricated and gives an array of approximately $10\lambda \times 10\lambda$ (or approximately $40 \times 40$~cm$^2$). A close view of each surface is found in Figure~\ref{fig_fab_surf}.

\section{Measurement methods}
\begin{figure}[]
		\centering
		\includegraphics[width=\linewidth]{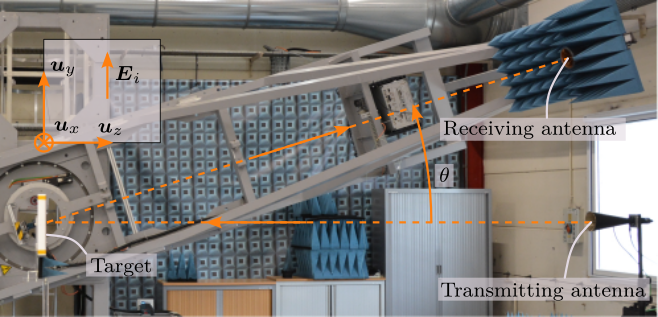}
		\caption{RCS measurement facility. A transmitting antenna at fixed position is illuminating the target while the bistatic $\theta$ scanning is performed by the receiving antenna, mounted on a rotating arm. The $\boldsymbol{E}$ field polarisation is shown in the top left corner.}\label{fig_under_test}
	\end{figure}

	The RCS measurements have been made with a specific 3D measurement facility at the CEA-CESTA, presented in~\cite{Massaloux2014}. The transmitting antenna is fixed and illuminates the metasurface under normal incidence, with a TM ($\boldsymbol{E}$ is contained in the reflection plane) polarisation. The receiving antenna is mounted on a rotating axis, also receiving in TM polarisation. The situation is depicted in Figure~\ref{fig_under_test} where both antennas are at a distance $r = 4$~m from the metasurface.
	
	For obvious obstruction reasons, observation angles $\theta$ for which $\lvert\theta\rvert < 10^\circ$are not accessible. This is not an issue since the observation angle of interest is $\theta = −60^\circ$. The facility in this configuration then gives access to angles from $10^\circ$ to $90^\circ$. The metasurface was flipped after the first angular sweep to reach negative observation angles. The calibrations are made with measurements on a canonical target, namely a standard metallic sphere with known analytical RCS, in the same conditions as with metasurface measurements. Even though the optimisation has been performed at a unique frequency, measurements are made over a 2~GHz to 18~GHz frequency range to apply spatial filtering. A classical time-gating approach is used.
	
	The wideband $S$ parameter measurements are exploited in the time (or distance) domain to remove the contribution of the antenna-target-room interaction from the received signal. An empty room measurement is also performed without any target, and the resulting $S$ parameter is subtracted from the metasurface $S$ parameter measurement. This procedure allows withdrawing the antenna-room interactions from the gathered signal. The abovementioned RCS measurement techniques allow us to avoid the use of great quantities of absorbing materials or special equipment and chambers. Only a few absorbing materials are used in the vicinity of the receiving antenna to avoid non direct paths that are close to the direct path, which would not be easy to filter out from raw data. Finally, RCS values in decibels are obtained taking the logarithm of the subtracted, calibrated and filtered $S$ parameters modulus.
	\begin{figure*}[h]
		\centering
		\includegraphics{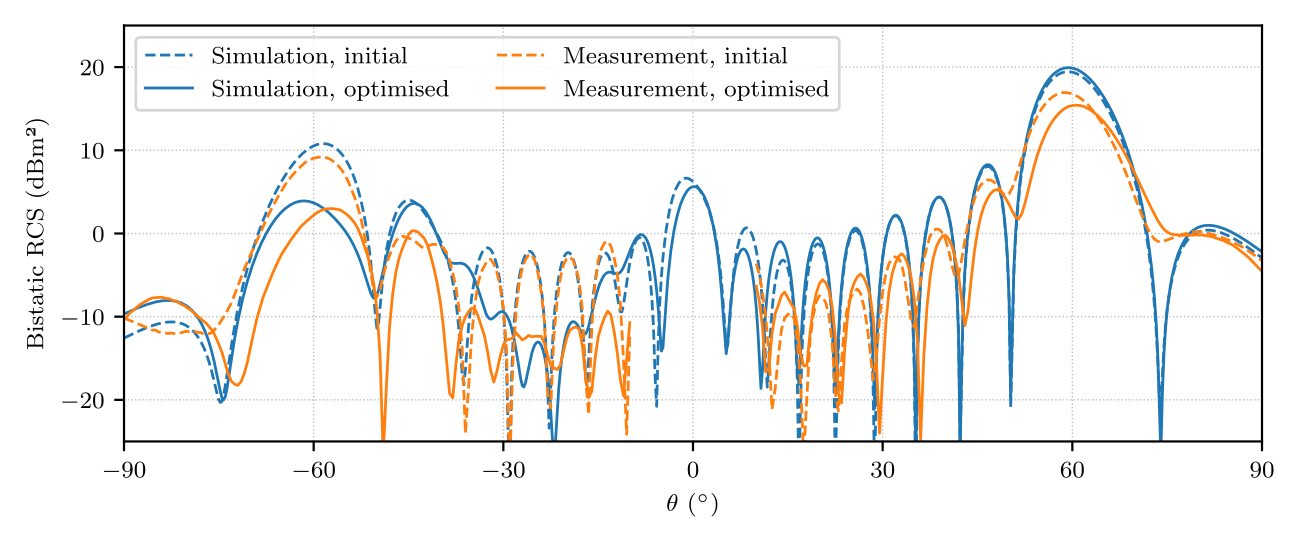}
		\caption{Bistatic RCS at $f = 8$~GHz of both initial and optimised metasurfaces, obtained  in full wave simulation with HFSS or with RCS measurement.}\label{fig_sim_mes}
	\end{figure*}

\section{Results}
	The measured RCS of the initial and optimised surfaces are represented in Figure~\ref{fig_sim_mes}, along with the simulated ones. Using the same comparison procedure as for simulations, the experimentally observed figure of merit at the target frequency of 8 GHz is $\delta\mathrm{RCS}=-7.7$~dB for the initial case and $\delta\mathrm{RCS}=-13.2$~dB for the optimised case, which is a $5.5$~dB measured relative reduction. The measured RCS in the $\theta = 60^\circ$ direction is also reduced by $1.5$~dB in the optimised case compared to the initial one. While this reduction is notable, in the context of stealth applications, the absolute RCS level in the $\theta = 60^\circ$ direction is of less importance. This direction corresponds to an area where it is possible to radiate but not mandatory. The primary objective is to minimize the lobe levels in directions where radiation is unintended which is, in this case, $\theta = −60^\circ$. Globally, the simulation overestimates RCS values. Actually, the number of assumptions for the synthesis of the surfaces is large (local periodicity, infinite environment, description with only a phase response, phase response computed at only one frequency). This large number of assumptions lightens the optimisation process to make it highly efficient. We can still discuss the two main hypothesis, the major one being the periodicity assumption which is discussed in the following paragraph.

	On the one hand, the main difference between simulation and measurement is the use of a finite surface in the latter case. The periodicity assumption then becomes less accurate as the cells are closer to an edge of a surface. On the other hand, it is clear that in the optimised case the cells and supercells are experiencing more coupling effect than in the initial case as they are closer to each other (see Figure~\ref{fig_fab_surf}). The two observations combined leads to the conclusion that, in the optimised case, the periodicity assumption is weaker than in the initial case, the cells being more dependent on their neighbors. This is indeed observed in Figure~\ref{fig_sim_mes}, where a less accurate overall correlation is found between simulation and measurement for the optimised case.
	
	The second hypothesis is the far-field assumption. From a measurement point of view, with a square metasurface measuring approximately $40$~cm on a side and a 4 meter radar range, the far field hypothesis is not accurate. The phase difference between the center and an edge of the metasurface in our configuration is of approximately $180^\circ$. In contrast, the phase response variations with the dimension $L$ of a cell as observed in~\cite{Elineau2022b} is of $280$~deg.mm$^{-1}$ at its maximum. For a fabrication tolerance of $50$~µm this gives, in the worst case, an uncertainty of $14^\circ$ which is way under the calculated phase difference of $180^\circ$.
	
\begin{figure}
		\centering
		\includegraphics{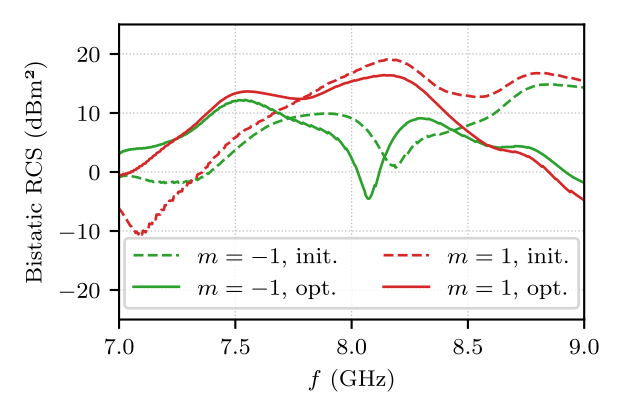}
		\caption{Evolution of the RCS over a frequency range centered around the optimisation frequency, in the $\theta_m$ directions, for the initial and optimised surface.}\label{fig_perf}
	\end{figure}

	A wideband study is proposed in order to further discuss the domain of validity of the optimisation. This study still focuses on RCS levels of the parasitic radiation direction relatively to the main radiation direction. Nevertheless, these two particular directions vary with frequency and are $\pm 60^\circ$ only at $8$~GHz. From Floquet theory they are, in the general case, $\theta_m=\arcsin⁡(\frac{m\lambda}{3d})$ with $m = \pm 1$. Figure~\ref{fig_perf} shows the frequency evolution of the RCS levels in the $\theta_{\pm 1}$ direction for initial and optimised structures. 
	For clarity purposes the trace domain is restricted to the $7$~GHz to $9$~GHz frequency band, while the measurement was, as said earlier, conducted in the $2$~GHz to $18$~GHz band. Around the central frequency, the distances between the two solid lines is increased in comparison to the distances between the two dashed lines. This is essentially the optimisation role to increase that distance. A similar explanation is that the level in the $\theta_1$ direction stays stable through the optimisation while the level in the $\theta_{-1}$ direction is significantly decreased. It is also observed that the parasitic lobe level local minimum is decreased and is getting closer to central frequency (from dashed to plain green line, from $8.2$ to $8.1$~GHz).
	
\begin{figure}
		\centering
		\includegraphics{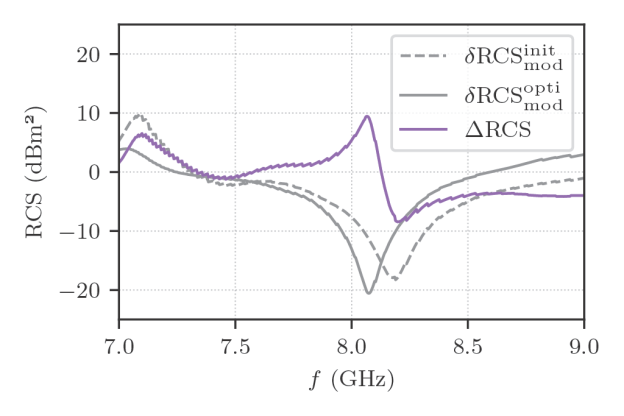}
		\caption{Frequency evolution of the modal figure of merit $\delta\mathrm{RCS}_\mathrm{mod}$ for the initial and optimised surface and of the difference $\Delta\mathrm{RCS}$ around the optimisation frequency.}\label{fig_f_perf}
	\end{figure}

	In a similar way to the first results discussions, a modal figure of merit $\delta\mathrm{RCS}_\mathrm{mod} = \frac{\mathrm{RCS}(\theta_{-1})}{\mathrm{RCS}(\theta_1)}$ is proposed. Its evolution for both the initial structure and the optimised structure is presented in grey lines in Figure~\ref{fig_f_perf}.
	The optimisation is considered efficient when the difference between the two figures ${\Delta\mathrm{RCS} = \delta\mathrm{RCS}_\mathrm{mod}^\mathrm{init} - \delta\mathrm{RCS}_\mathrm{mod}^\mathrm{opti}}$ is positive. This quantity, which ultimately describes the wideband parasitic lobe level decrease relatively to the main lobe level, is also presented in Figure~\ref{fig_f_perf} in purple line. Similarly to Figure~\ref{fig_perf} observations, the local minimum of $\delta\mathrm{RCS}_\mathrm{mod}$ is diminished and shifted towards central frequency during the optimisation procedure. This representation also highlights the narrowband behaviour of the optimisation, as $\Delta\mathrm{RCS}$ is positive only over a few hundreds of MHz around the central frequency. Indeed, no design rule has been followed toward the synthesis of a wide band device. If any wideband performance is required, the supercell optimisation has to be performed over the desired frequency band. More generally, metasurfaces devices fail to handle wideband phenomena, or at the cost of sophisticated synthesis procedures (multi scales or multi-layer devices~\cite{Samadi2021}) that are not practical for stealth applications.

\section{Conclusion}
	This study demonstrates the efficiency of the RCS reduction metasurface optimisation method by proposing a measurement of such a surface. The simulation and measurement exhibit a RCS reduction of almost $7$~dB, in the desired direction, with an excellent agreement between them. It has been made possible by the simplicity of the Floquet type simulation and the correct use of the outputs of these simulations to predict the behavior of the surface. The method should not be limited to the RCS reduction in a particular direction. One could consider applying this method for any Floquet mode distribution, addressing one or multiple modes. The literature provides various recent analyses of multichannel Floquet metasurfaces (in~\cite{Vuyyuru2024} for example), but none of them approaches the problem from a RCS point of view. We think that this end to end study builds the bridge between Floquet analysis and RCS estimation. Multichannel RCS reduction should not show more limitations than the ones already coming from classical multichannel Floquet metasurfaces.

\section*{Acknowledgments}
	The authors thank Guillaume Cartesi and Olivier Raphel for their great contribution during the measurement campaign.

\printbibliography%

\end{document}